\documentclass[structabstract]{aa}  
\usepackage{graphicx}
\usepackage{txfonts}

\usepackage{natbib}
\bibpunct{(}{)}{;}{a}{}{,}

\def\depthsep{0.244$^{+0.027}_{-0.020}$ }

\def\TBsep{3270$^{+115}_{-160}$ }

\def\sigsep{12 }

\begin{document}
\title{The GROUSE project III: Ks-band observations of the thermal emission from WASP-33b\thanks{Photometric timeseries are only available in electronic form at the CDS via anonymous ftp to cdsarc.u-strasbg.fr (130.79.128.5) or via http://cdsweb.u-strasbg.fr/cgi-bin/qcat?J/A+A/}}
   \titlerunning{The GROUSE project III: The secondary eclipse of WASP-33b}
   \authorrunning{De Mooij et al.}
   \author{E.J.W. de Mooij\inst{1,2}, 
           M. Brogi\inst{1},
           R.J. de Kok\inst{3},
           I.A.G. Snellen\inst{1},
           M.A. Kenworthy\inst{1},
           \and
           R. Karjalainen\inst{4}}
   \institute{Leiden Observatory, Leiden University, Postbus 9513, 2300 RA, Leiden, 
               The Netherlands;
 \and Department of Astronomy and Astrophysics, University of Toronto, 50 St. George Street, Toronto, ON M5S 3H4, Canada;    \email{demooij@astro.utoronto.ca}
             \and SRON Netherlands Institute for Space Research, Sorbonnelaan 2, 3584 CA Utrecht, The Netherlands;   
             \and Isaac Newton Group of Telescopes, Apartado de Correos 321, E-38700 Santa Cruz de la Palma, Canary Islands, Spain}

\abstract
{In recent years, day-side emission from about a dozen hot Jupiters has been detected through ground-based secondary eclipse observations in the near-infrared. These near-infrared observations are vital for determining the energy budgets of hot Jupiters, since they probe the planet's spectral energy distribution near its peak.}
{The aim of this work is to measure the K$_s$-band secondary eclipse depth of \mbox{WASP-33b}, the first planet discovered to transit an A-type star. This planet receives the highest level of irradiation of all transiting planets discovered to date. Furthermore, its host-star shows pulsations and is classified as a low-amplitude $\delta$~Scuti.}
{As part of our GROUnd-based Secondary Eclipse (GROUSE) project we have obtained observations of two separate secondary eclipses of \mbox{WASP-33b} in the K$_s$-band using the LIRIS instrument on the William Herschel Telescope (WHT). The telescope was significantly defocused to avoid saturation of the detector for this bright star (K$\sim$7.5). To increase the stability and the cadence of the observations, they were performed in staring mode. We collected a total of 5100 and 6900 frames for the first and the second night respectively, both with an average cadence of 3.3 seconds.}
{On the second night the eclipse is detected at the  \sigsep-$\sigma$ level, with a measured eclipse depth of \depthsep\%. This eclipse depth corresponds to a brightness temperature of \TBsep~K. The measured brightness temperature on the second night is consistent with the expected equilibrium temperature for a planet with a very low albedo and a rapid re-radiation of the absorbed stellar light. For the other night the short out-of-eclipse baseline prevents good corrections for the stellar pulsations and systematic effects, which makes this dataset unreliable for eclipse depth measurements. This demonstrates the need of getting a sufficient out-of-eclipse baseline.}
{}


\maketitle

\section{Introduction} 
In recent years, there have been many measurements of thermal emission from the atmospheres of hot Jupiters, especially in the mid-infrared using the {\it Spitzer} Space Telescope \citep[e.g. the review by][]{deming09iau}. These {\it Spitzer} observations probe the thermal emission of hot Jupiters redward of the peak of their spectral energy distribution (SED), and thus measure the planet's light in the Rayleigh-Jeans tail of their emission spectrum. 

Observations in the near-infrared, on the other hand, typically probe the planet's emission spectrum around or even blue-ward of its peak, and therefore provide essential information on the planet's total energy budget. During the past three years several measurements of planetary emission shortward of 2.5$\mu$m have been obtained \citep[][]{demooijandsnellen09, singandlopezmorales09, gillonetal09,rogersetal09, andersonetal10, alonsoetal10, gibsonetal10, crolletal10a, lopezmoralesetal10, crolletal10b, crolletal10c, demooijetal11, smithetal11, caceresetal11,demingetal12},  most of these are in the K-band ($\lambda$=2.15$\mu$m) where the planet-to-star contrast is most-favourable for observations with ground-based telescopes through the available atmospheric windows.

From the combination of the measurements at multiple wavelengths, a picture is emerging that there are (at least) two types of hot Jupiter atmospheres, those which show a thermal inversion, and those which do not. It has been proposed that the presence of the inversion layer is set by the stellar irradiation, where at high levels of irradiation the planet's stratosphere is hot enough to keep a strongly absorbing compound in the gas phase, while at lower irradiation levels the compound condenses out and disappears from the gas-phase \citep[e.g.][]{burrowsetal07,fortneyetal08}. 

\cite{knutsonetal10} proposed an alternative scenario for the presence or absence of a strong absorber in the highest layers of the planetary atmosphere. In their scenario the absorber can be destroyed by strong UV emission from the planet's host-star, due to stellar activity. For higher levels of stellar activity, which result in a higher UV flux, the absorbing compound is possibly removed, resulting in a non-inverted atmosphere. Note that the inference of an inversion layer has recently been questioned by \cite{madhusudhanandseager10}, who point out that for several planets there is a degeneracy between the atmospheric temperature structure and the chemical composition of the planet's atmosphere.

In this paper we present observations of two secondary eclipses of the very hot Jupiter \mbox{WASP-33b} in K$_s$-band. These are part of the GROUnd-based Secondary Eclipse project (GROUSE), which aims to use ground-based telescopes for exoplanet secondary eclipse observations in the optical and near-infrared. As part of this project we have already published K$_s$-band detections of the secondary eclipses of \mbox{TrES-3b} \citep[][]{demooijandsnellen09} and \mbox{HAT-P-1b} \citep[][]{demooijetal11}

\mbox{WASP-33b}  \citep{colliercameronetal10} is currently the only known planet to transit an A-type star (T$_{\rm{eff}}$=7430$\pm$100K) which it orbits in $\sim$1.22 days. This makes \mbox{WASP-33b} the most irradiated planet known to date, with an irradiation of 1.2$\cdot$10$^{10}$ erg/sec/cm$^2$. This high level irradiation results in an expected day-side equilibrium temperature of 3250K.  Recent observations by \cite{smithetal11} indeed show a very high brightness temperature at 0.9$\mu$m of 3466$\pm$140K. Additionally, since the host-star is relatively hot, the expected UV flux it receives is also high, making it an ideal candidate to investigate the influence of a high UV flux on the temperature structure of a planet's atmosphere.

In addition to being the first transiting planet discovered to orbit an A-type star, \mbox{WASP-33b} is also the first planet to transit a pulsating star. In the discovery paper, \cite{colliercameronetal10} find evidence for non-radial pulsations in their spectral time-series, and tentatively classified \mbox{WASP-33} as a $\gamma$~Doradus pulsator, which is a class of non-radial pulsators with periods of $\sim$0.3 days or longer \citep[see e.g.][]{handlerandshobbrook02}. Recently, \cite{herreroetal11} analysed photometric time series for \mbox{WASP-33} and found a pulsation period of 68.6 minutes, which, when converted to the pulsation parameter Q, the product of the pulsation period and the square-root of the mean stellar density \citep[e.g.][]{breger90,handlerandshobbrook02}, is comparable to that of $\delta$~Scuti stars, and well outside the range of $\gamma$~Doradus stars. The observed stellar pulsations have a measurable impact on the transit and eclipse measurements for this planet~\citep[e.g.][]{herreroetal11,smithetal11,demingetal12}.

In Sect.~\ref{sec:obs_dr} we present our observations and data reduction. In Sect.~\ref{sec:cor} stellar pulsations and the light curve fitting are presented. Subsequently we discuss the results in Sect.~\ref{sec:res}, and finally we will give our conclusions in Sect.~\ref{sec:concl}.

\section{Observations and data reduction}\label{sec:obs_dr}
\begin{figure}
\centering
\includegraphics[width=8cm]{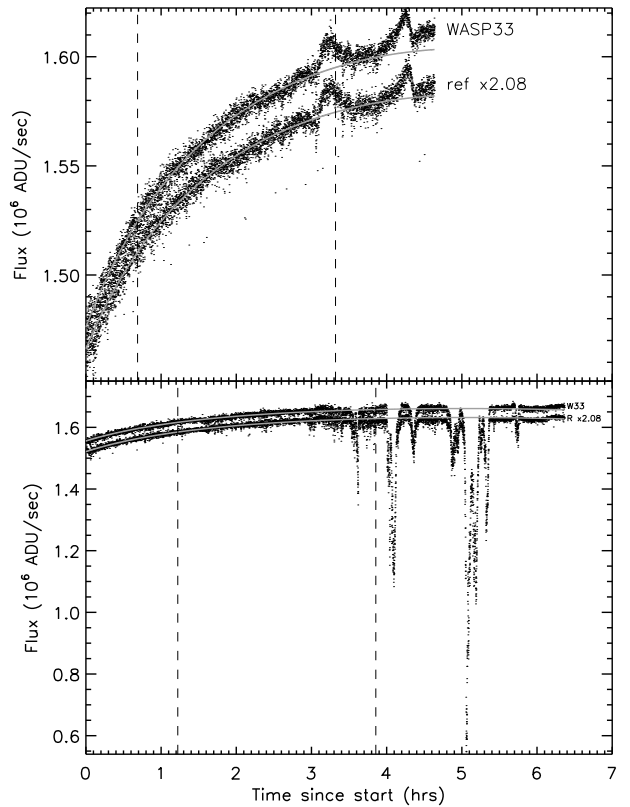}
\caption{Raw lighcurves for \mbox{WASP-33} and the reference star (multiplied by 2.08 for plotting purposes) for the night of August 18, 2010 (top panel) and for the night of September 20, 2010 (bottom panel). The vertical dashed lines indicate the expected beginning and end of the targeted eclipse. The solid, grey, lines show the airmass during the nights, scaled to match the stellar flux in the first hour of the observations.}
\label{fig:raw_lc}
\end{figure}

\subsection{Observations}
The  secondary eclipse of \mbox{WASP-33b} was observed on two nights, on August 18, 2010 and September 20, 2010, in the K$_s$-band with the Long-slit Intermediate Resolution Infrared Spectrograph \cite[LIRIS;][]{LIRIS02} instrument on the  William Herschel Telescope (WHT) on La Palma.

The pixel scale of LIRIS is 0.25 arcsec per pixel, yielding a field-of-view of 4.2 by 4.2 arcminutes, large enough to observe both \mbox{WASP-33} and a reference star of similar brightness simultaneously. Since \mbox{WASP-33} is very bright, exposure times of 1.5 seconds were used in order to avoid saturation of the detector. As an additional measure to prevent saturation, the telescope was strongly defocused. This is a well proven strategy also used for other GROUSE observations \citep[][]{demooijandsnellen09, demooijetal11}, which should also reduce the impact of flat-field inaccuracies by spreading the light over many pixels, thereby minimizing the impact from uncorrected pixel-to-pixel sensitivity variations. To keep the observations as stable as possible, and in order to reduce the cycle time, the observations were performed in staring mode. Since this method does not allow for background subtraction using the science frames, a set of sky frames were obtained after the observations on both nights for sky-subtraction purposes. 

On August 18 (Night I) the observations started at 00:45 UT and lasted for $\sim$4.5 hours. The weather conditions during the night were photometric, as can be seen from the raw light curves shown in the top panel of Fig.~\ref{fig:raw_lc}. A total of 5100 science frames were obtained with an average cadence of 3.3 seconds. The first three frames of a sequence of frames\footnote{The observations consisted of multiple sequences of 100-200 frames} are known to suffer from the reset anomaly, which is seen as an anomalous structure in the background. These frames are therefore excluded from further analysis, which results in a total of 4994 frames.

The observations on September 20 (hereafter night II) were taken between 22:35 UT and 05:00UT. During the first part of the observations the conditions were photometric, however during the last few hours occasional clouds moved across the image, absorbing up to 65\% of the light (see Fig.~\ref{fig:raw_lc}).  A total of 6693 science frames were obtained with an average cadence of 3.3 seconds per frame, excluding 207 frames due to the reset-anomaly.

\subsection{Data reduction}
The data-reduction for both nights was performed in the same way. All frames were corrected for crosstalk along rows of the detector, which is present at a level of 10$^{-5}$ of the total flux along the rows of all four quadrants. Subsequently we performed a non-linearity correction on all the frames using our own non-linearity measurements which were created from a set of dome-flats at a constant level of illumination but with varying exposure times. After these corrections the images were flat-fielded using a flat-field created from bright and dark twilight flats.

A background map was constructed from the set of dithered images obtained after the eclipse observations. These images were reduced in the same way as the science images, and, after filtering out the discrepant pixels in time to remove (faint) stars, were subsequently combined. The resultant background map was then scaled and subtracted from the individual science images.

After background subtraction, aperture photometry was performed on both \mbox{WASP-33} and the reference star using an aperture of 18 and 26 pixels for night I and night II respectively. Any residual sky background was determined in annuli between 30 and 50 pixels for night I and between 40 and 60 pixels for night II. The flux in the annuli was clipped at 5$\sigma$ to avoid outliers in the background (such as hot pixels) from affecting the data. Finally, the light curve of \mbox{WASP-33} was normalised with that of the reference star, and the resultant light curves for the two nights are shown in the top panels of Figs.~\ref{fig:lc_poly} and~\ref{fig:lc_sys}.

\section{Correction for systematic effects and stellar pulsations}\label{sec:cor}
\begin{figure}
\centering
\includegraphics[width=8cm]{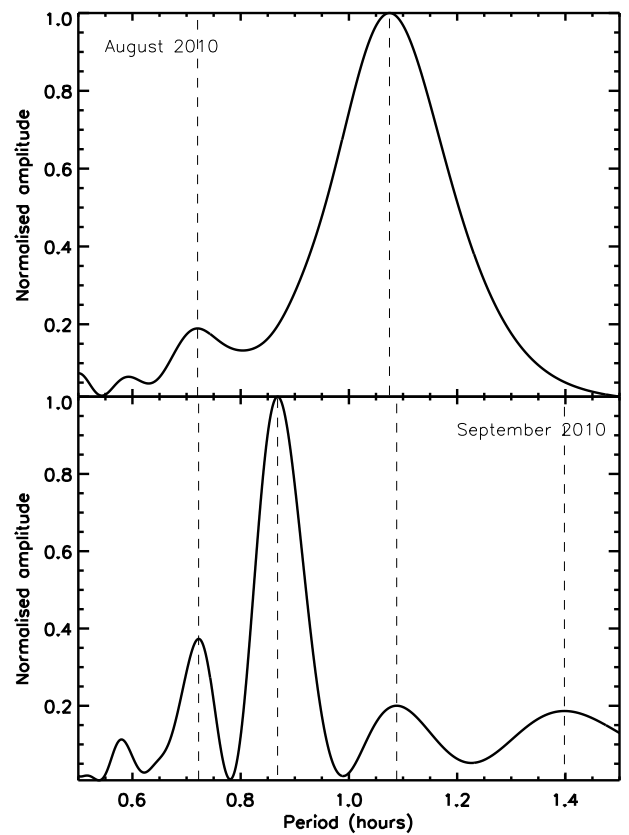}
\caption{Normalised periodograms of the light curves for the two separate nights. The top panel shows the periodogram for night I and the bottom panel shows the periodogram for night II. The dashed lines in both panels indicate the periods used in the fitting.}
\label{fig:periodogram}
\end{figure}

\subsection{Stellar pulsations}
\cite{colliercameronetal10} noted that the host-star of \mbox{WASP-33b} is a pulsator. Observations by \cite{herreroetal11} indicated a dominant period of 68.5 minutes. \cite{smithetal11} observed a secondary eclipse of \mbox{WASP-33b} in a narrowband filter at 0.91$\mu$m, and found three pulsation periods in their data, at 53.62, 76.52 and 41.85 minutes, all with amplitude between 0.4 mmag and 0.9 mmag. Observations by \cite{demingetal12} showed various pulsation periods, 68 minutes for observations of two separate events in the mid-infrared from the {\it Spitzer} Space Telescope, 71 minutes for one ground-based event in the K-band, 146 minutes for observations in the J-band and 54 \& 126 minutes for a second set of K-band observations. For all their periods the amplitudes were larger than 1 mmag, with the two nights of K-band data showing amplitudes in excess of 2 mmag.

For our data, the stellar pulsations are clearly visible in the light curve for night I, while for night 2 the variability is less apparent. In order to determine the period(s) of the stellar pulsations, a periodogram of the light curves was created. Since the planetary eclipse signal and possible systematic effects can influence the periods found in the data, both a scaled eclipse model as well as a model for the systematics based on instrumental effects (see section~\ref{sec:eclfit}) were fitted to the data using a simple linear regression algorithm and subsequently divided out before determining the periodogram. The generalised Lomb-Scargle formalism from \cite{zechmeisterandkurster09} was used to construct the periodogram, and the results for both nights are shown in Fig.~\ref{fig:periodogram}.

For night I the strongest peak in the periodogram is found at a period of 64.5 minutes, and there is a weaker peak at 43.2 minutes. For night II, there are four peaks visible, the strongest peak is found at 52.1 minutes, while three weaker peaks are found at 43.3, 65.3 and 83.9 minutes. The signals at all the periods in both datasets have a false-alarm probability, estimated using Monte-Carlo simulations, of below 0.1\%, indicating that the periodicities are very likely real, although they might not be astrophysical in origin.
 In both datasets we find a periodic signal at $\sim$65 minutes, which differs from the period of 68.5 minutes found by \cite{herreroetal11} and the 68 minute period from the two {\it Spitzer} datasets from \cite{demingetal12}. However, the short time span covered  during each night is not sufficient to get a very tight constraint on the period, and therefore the periods could be consistent with 68 minutes. 
The $\sim$43 minute period is seen in the measurements from both our nights, as well as in the data from \cite{smithetal11}. The period around $\sim$52 minutes is found in both our second night of data as well as in the data from \cite{smithetal11} and \cite{demingetal12}, although there it is not the dominant frequency. We caution, however, that the periodograms used for the frequency analysis were created with data that was only partially corrected for systematic effects, and therefore can still be influenced by residual (quasi) periodic systematic effects.

\subsection{Light curve fitting}\label{sec:eclfit}

\begin{table*}
\caption{Fitted parameters and uncertainties from the MCMC analysis of the lightcurves of WASP-33b. }
\label{tab:fittab}
\centering
\renewcommand{\arraystretch}{1.7}
\begin{tabular}{c c c c c c}
\hline\hline
\multicolumn{2}{c}{Parameter}     &     \multicolumn{2}{c}{Night  1}    &   \multicolumn{2}{c}{Night 2} \\
     &    unit       &  Instrumental parameters & Polynomial & Instrumental parameters & Polynomial  \\
\hline
F$_p$/F$_*$   &   (\%)   &   0.140$\pm$0.007  &   0.092$\pm$0.017  &   0.245$\pm$0.009  &   0.246$\pm$0.018  \\  
\hline
x   &    &   0.06$\pm$0.01  &   ---   &    -0.12$\pm$0.02   &  ---   \\   
airmass   &    &  -0.76$\pm$0.03  &   ---   &     0.34$\pm$0.03   &  ---   \\  
sky   &    &   0.27$\pm$0.03  &   ---   &    -0.53$\pm$0.03   &  ---   \\   
c$_1$   &   &  ---   &     0.42$\pm$0.01  &   ---   &     0.10$\pm$0.01   \\  
c$_2$   &   &  ---   &    -0.09$\pm$0.02  &   ---   &    -0.01$\pm$0.02   \\  
c$_3$   &   &  ---   &     0.10$\pm$0.01  &   ---   &    -0.13$\pm$0.01   \\  
\hline
P$_1$   &   (minutes) &  63.95$\pm$0.39   &    62.00$\pm$0.51  &    52.65$\pm$0.39   &    52.28$\pm$0.51   \\  
P$_2$   &   (minutes) & \multicolumn{2}{c}{ 43.22 (fixed)}   &   \multicolumn{2}{c}{ 43.34 (fixed)}   \\  
P$_3$   &   (minutes) & \multicolumn{2}{c}{---}    &   \multicolumn{2}{c}{ 65.29 (fixed)}   \\
P$_4$   &   (minutes) & \multicolumn{2}{c}{---}    &   \multicolumn{2}{c}{ 83.90 (fixed)}   \\
A$_1$   &   (\%)   & 0.095$\pm$0.004   &   0.082$\pm$0.005   &   0.056$\pm$0.006   &   0.058$\pm$0.006   \\  
A$_2$   &   (\%)   & 0.041$\pm$0.004   &   0.042$\pm$0.004   &   0.017$\pm$0.005   &   0.014$\pm$0.005   \\  
A$_3$   &   (\%)   & \multicolumn{2}{c}{---}  &   0.011$\pm$0.006   &   0.011$\pm$0.006   \\  
A$_4$   &   (\%)   & \multicolumn{2}{c}{---}  &   0.031$\pm$0.006   &   0.013$\pm$0.006   \\  
\hline
\hline
\end{tabular}
\tablefoot{Baseline coefficents for the instrumental model are the x-position, airmass and sky level and for the polynomial baseline the coefficients are c$_1$ to c$_3$. P$_1$ to P$_4$ are the periods used for the stellar pulsations, and A$_1$ to A$_4$ are the corresponding amplitudes.}
\end{table*}

\begin{figure*}
\centering
\includegraphics[width=16cm]{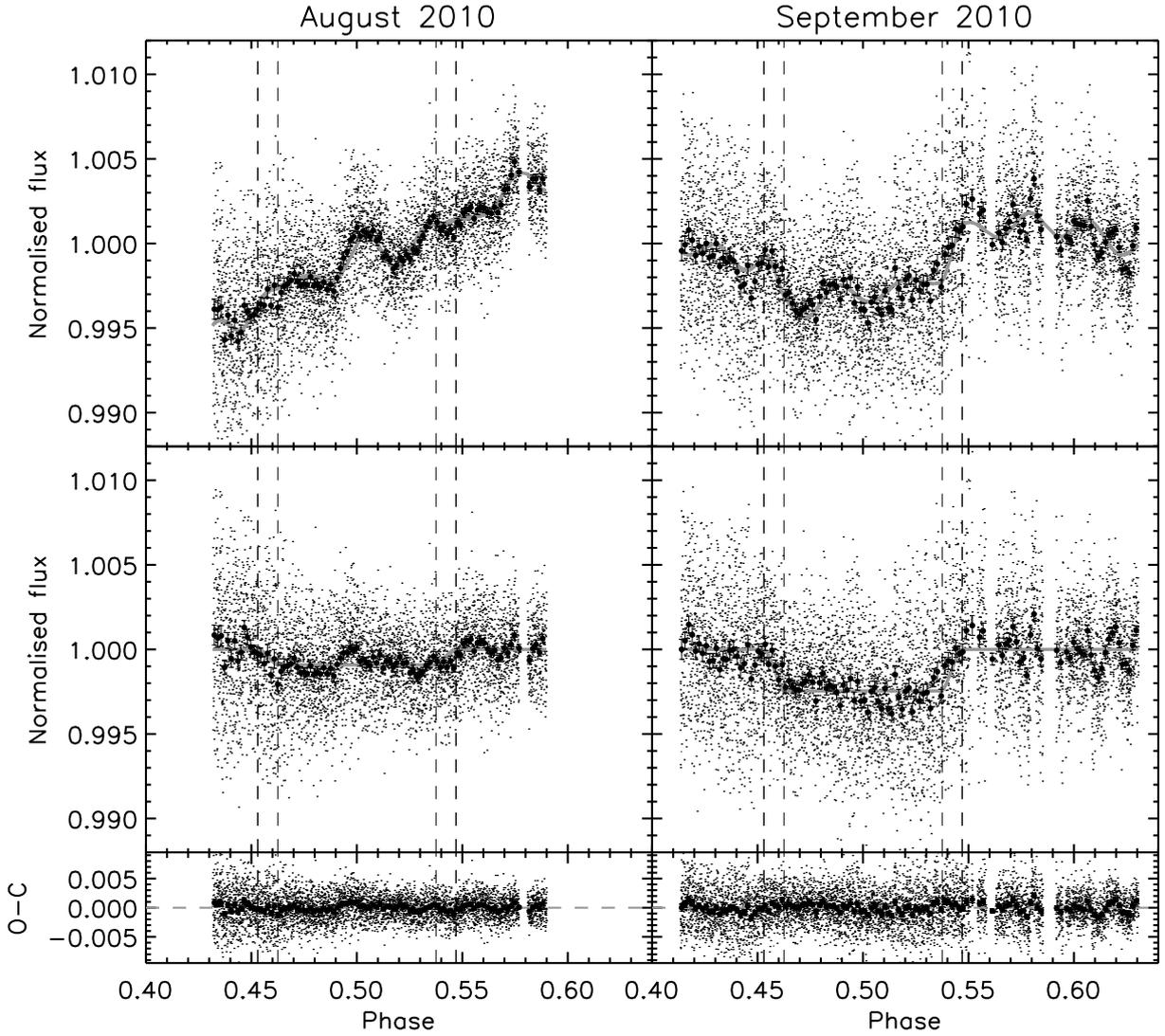}
\caption{Light curves for the secondary eclipse of \mbox{WASP-33b} for the night I (left panels) and night II (right panels). Top panels: the light curves of \mbox{WASP-33} normalised with those of the reference star, overplotted is the best fitting 'full' model with a low order polynomial baseline correction, stellar pulsations and the eclipse. Middle panels: The light curves corrected for the trends in the baseline and stellar pulsations, clearly showing the transit. Bottom panels: The residuals after subtracting the best-fit model. The thick points with errorbars in these figures show the data binned by 50 points. The vertical dashed lines show the expected times for first to fourth contact.}
\label{fig:lc_poly}
\end{figure*}

\begin{figure*}
\centering
\includegraphics[width=16cm]{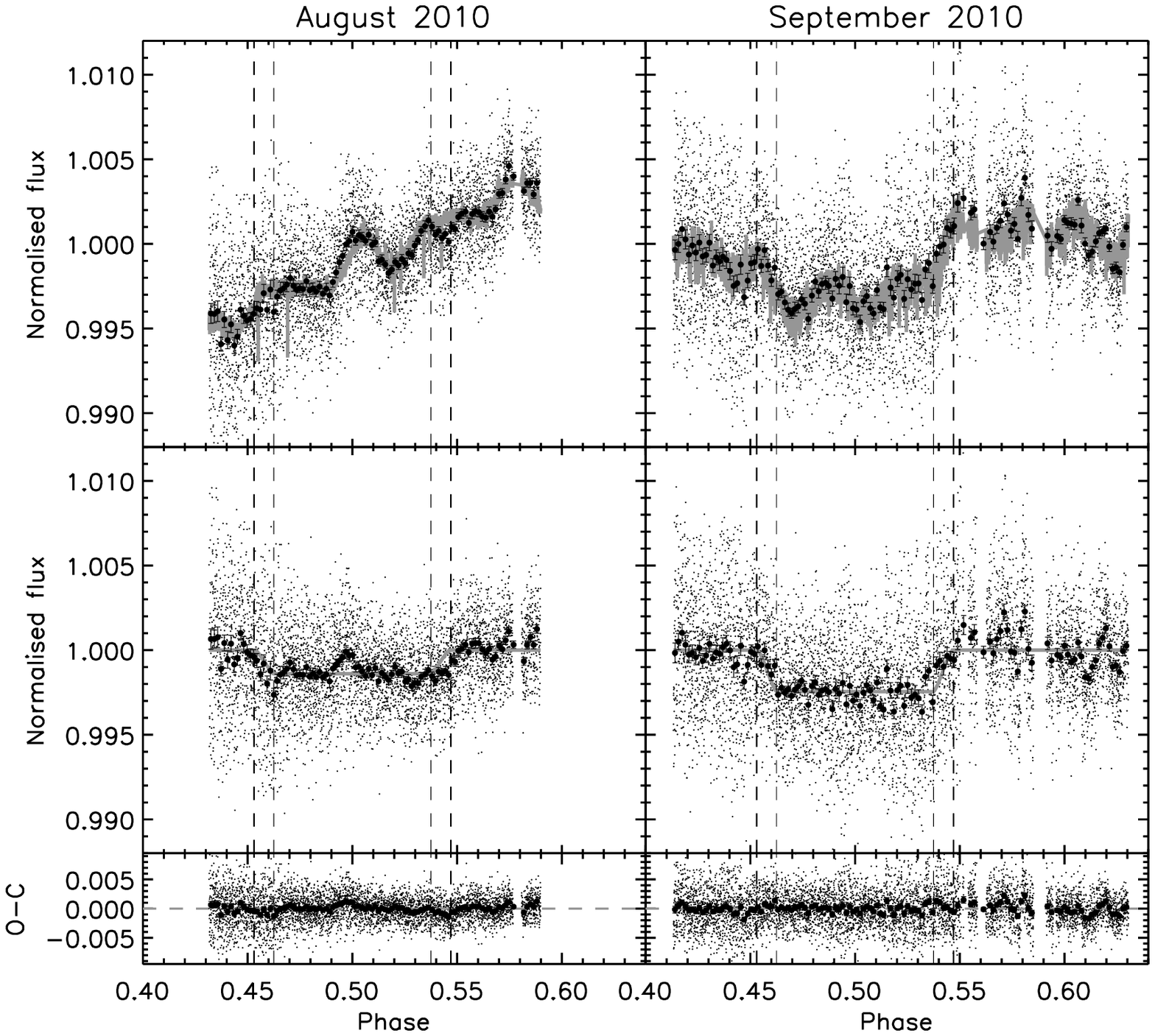}
\caption{Same as Fig.~\ref{fig:lc_poly} but the systematic effects are now modelled by linear relations with instrumental parameters.}
\label{fig:lc_sys}
\end{figure*}

Since the light curve is the result of a combination of three effects, the stellar variability, systematic effects related to both the instrument and the Earth's atmosphere and the secondary eclipse of \mbox{WASP-33b}, a fit for all three effects is performed simultaneously. 

For the stellar pulsations the period of the dominant mode is left as a free parameter, although with a penalty for the $\chi^2$ of the form (P-P$_0$)$^2$/$\sigma_p^2$, with P$_0$ the period determined from the periodogram, and $\sigma_p$ set to 1 minute. The periods of the other modes was kept fixed to periods found in the analysis of the periodogram as described in the previous section. For all the modes the offset in phase and the amplitude of the pulsations were allowed to vary freely. 

For the fitting of systematic effects two different methods were used. For the first method, the systematic effects are considered to be due to the change of position on the detector, the airmass and the difference in sky background between the two quadrants. This is similar to what was used in the previous papers from the GROUSE project \citep[][]{demooijandsnellen09, demooijetal11}. In the second method, the systematic effects are modelled using low order polynomials, as also used in \cite{demooijetal12} for the near-infrared transit observations of GJ1214b.  

The secondary eclipse was modelled using the \cite{mandelandagol02} formalism. We used the parameters from \cite{colliercameronetal10} for the impact parameter, semi-major axis, orbital period and planet-to-star size ratio, while the orbit of the planet is assumed to be circular. This assumption is reasonable since the planet orbits extremely close to its host-star, which should  result into a rapid damping of eccentricity. In addition, on the second night an eclipse-shaped dip in the light curve centered on $\phi\sim$0.5 is readily visible (see the right panel of Fig.~\ref{fig:lc_poly}). 

Before fitting, outliers were removed by excluding all points that were more than 0.9\% away from a median smoothed light curve with a box size of 51 points, as well as all points for which the flux of the individual stars, corrected for airmass, dropped below 90\%. In this way a total 21 and 414 points were excluded during the first and second night respectively. In addition, there is a feature present in both  light curves at the same time after the start of the observations (after 0.1795$\pm$0.0025 days) that is not at an identical point during the planet's orbit and therefore most likely due to an, as yet, unidentified instrumental effect. Excluding all the points that were obtained during this feature removes an additional 128 frames.

The light curves were fitted with 9 free parameters (1 for the eclipse, 3 for the systematic effects and 5 for the stellar pulsations) for night I, and 13 free parameters for night II (due to 2 additional periods found in the data).  The two nights were fitted separately using a Markov-Chain Monte Carlo method. Per night, 5 sequences of 2 million steps were generated, trimming the first 200,000 points to avoid any contamination from the initial conditions. The chains were combined after checking that they were well mixed \citep{gelmanandrubin92}.

\section{Results and discussion}\label{sec:res}

The best fit values of the eclipse depth  and their fornal uncertainties for night I are 0.140$\pm$0.007\% and 0.092$\pm$0.017\% for the fit with instrumental parameters and polynomials respectively, while for night II the best-fit eclipse depths are 0.245$\pm$0.009\% and 0.245$\pm$0.018\% for the two respective anlyses. These results are given in Table~\ref{tab:fittab}.

The differences found between methods in the first night are significantly larger than the uncertainties in the eclipse depth as estimated from the MCMC analysis. First of all the first night suffers from a strong peak, possibly due to stellar pulsations, right in the middle of the eclipse. We attribute this to a problem with the observations on night I. The relatively short out-of-eclipse baselines available for the first eclipse observation hampers the removal of the systematic effects from the stellar pulsations, as well as from the instrumental and atmospheric effects. This is clearly illustrated when looking at the correlations between the parameters used for the removal of the systematics and the eclipse depth, as shown in Figs.~\ref{fig:mcmc_cor_n1pol} to~\ref{fig:mcmc_cor_n2sys}. We therefore conclude that the first night of data is no usable  for a reliable eclipse measurement.

To assess the impact of correlated noise, we redid the analysis for night II after binning the data by 50 points ($\sim$3 minutes). Although overall the parameters are the same, we find larger uncertainties in the eclipse depths with the best-fit values of 0.255$\pm$0.028\% and 0.242$\pm$0.035\% for the fit with instrumental parameters and polynomials respectively. To assess the impact of (uncorrected) red noise on the measured eclipse depths in another way, the residual permutation method was used \citep[e.g.][]{gillonetal07}. The best fit model is subtracted from the light curve, and these residuals are then shifted by n points, wrapping the light curve around, so that the points that are shifted beyond the end of the lightcurve are inserted at the beginning. The best fit model is then added back to the data, and this new light curve is fitted again. The interval between  16\% and 84\% of the distribution of the best-fitting eclipse depths is used for the 1-$\sigma$ uncertainties on the eclipse depth. To speed up the residual permutation analysis, instead of adding back the full model, which includes the stellar pulsation, trends in the baseline and eclipse depth, we only used the trends in the baseline and eclipse depth, since the correlation between the parameters for the stellar pulsations and the eclipse depth is weak. 
From the residual permutation analysis we also find larger uncertainties for both decorrelation methods, with eclipse depths of 0.244$^{+0.027}_{-0.020}$\% for a baseline fitted with instrumental parameters and 0.249$^{+0.033}_{-0.052}$\% for a polynomial baseline fit. In all cases the uncertainties are higher than for the MCMC analysis of the unbinned data but comparable to the MCMC analysis of the binned data, which is expected in the presence of red noise.

\begin{figure}
\centering
\includegraphics[width=8cm]{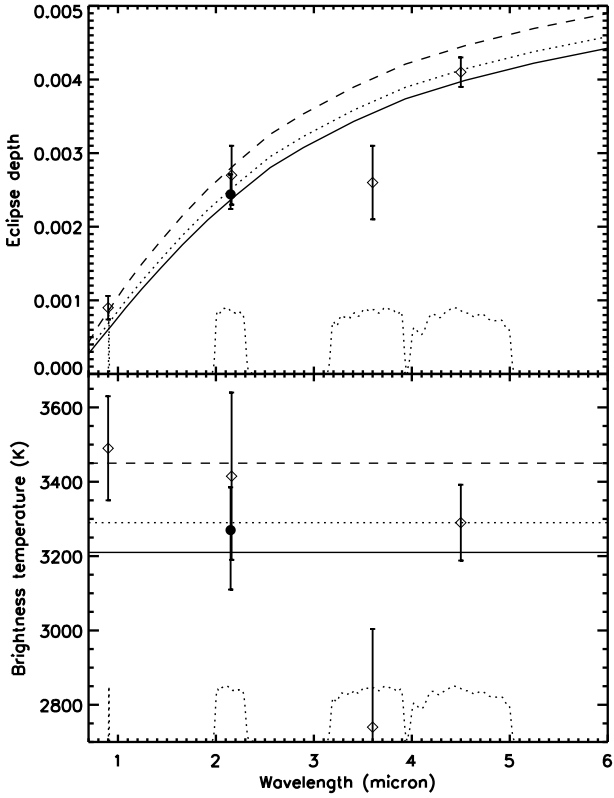}
\caption{Spectral energy distribution of \mbox{WASP-33b}. Top panel: Eclipse depths in K$_s$-band (this work) and in the SII$_{0.91\mu m}$-filter from \cite{smithetal11}. Bottom panel: brightness temperatures in the two bands. Overplotted in both panels are the expected eclipse depths/brightness temperatures for a zero-albedo homogeneous day-side (solid line), for an instantly re-radiating day-side (dashed line) and for the best-fit effective temperature (dotted line).}
\label{fig:sed}
\end{figure}

Since there is a strong correlation between the coefficients for the polynomial baseline fit and the eclipse depth (see Fig.~\ref{fig:mcmc_cor_n2pol}), we use the fit of the baseline with instrumental parameters for the remainder of the paper, since the correlation between different parameters is much weaker. We note that the polynomial baseline correction for this night gives the same eclipse depth, however with a larger uncertainty.

The measured eclipse depth of \depthsep \% corresponds to a brightness temperature in the K$_s$-band of  \TBsep K. This brightness temperature was calculated using the solar-metallicity NextGen models \citep{hauschildtetal99} interpolated to the stellar parameters of \mbox{WASP-33} determined by \cite{colliercameronetal10} (T$_{\rm{eff}}$=7430K, log(g)=4.294).

Currently there are several other measurments of the secondary eclipse of \mbox{WASP-33b}, \cite{smithetal11} obstained data in a narrowband filter at $\sim$9100\AA, showing a depth of 0.09$\pm$0.016\%. \cite{demingetal12} observed both in the Ks-band and in the IRAC 3.6$\mu$m and 4.5$\mu$m bands, obtaining eclipse depths of 0.27$\pm$0.04\%, 0.26$\pm$0.05\% and 0.41$\pm$0.02\% respecively. These brightness measurements correspond to a brightness temperature of 3490$\pm$140~K, 3415$\pm$130~K, 2740$\pm$225~K, 3290$\pm$100~K for the SII, K$_s$, IRAC 3.6$\mu$m and 4.5$\mu$m respectively. Although a planet's brightness temperature can be a strong function of wavelenght, most of the  measurements are consistent at the 1$\sigma$ level.

As can be seen in the middle right panel of Fig.~\ref{fig:lc_sys}, the eclipse appears to end earlier than expected from the model. Although systematic effects are the most likely cause of this, it is worth noting that a narrower width of the eclipse would indicate that the orbit of \mbox{WASP-33b} is eccentric. If this is the case, by combining the ratio between the transit and secondary eclipse durations with the time of mid-eclipse, a direct measurement of both the eccentricity and the argument of periastron is possible \citep[e.g.][]{charbonneauetal05}. Although a full fit is beyond the scope of this work, we can estimate the change in duration from the light curve. The duration of the secondary eclipse is shorter than the transit duration by 0.01 in phase, such that the eclipse duration corresponds to $\sim$90\% of transit duration. From this we estimate esin($\omega$)$\sim$0.05. Since the ingress appears to be at the expected time, the time of mid-eclipse is in this case also slightly earlier than expected. From the shift we  estimate ecos($\omega$)$\sim$0.008. Combining these two estimates we find an eccentricity of $\sim$0.05. We again caution that systematics can easily give rise to an apparent non-zero determination of the eccentricity, which is for instance seen for the secondary eclipse of TrES-3b \citep[][]{demooijandsnellen09,fressinetal10,crolletal10b}. We therefore do not advocate this non-zero eccentricity scenario based on these data.

\begin{figure}
\centering
\includegraphics[width=8cm]{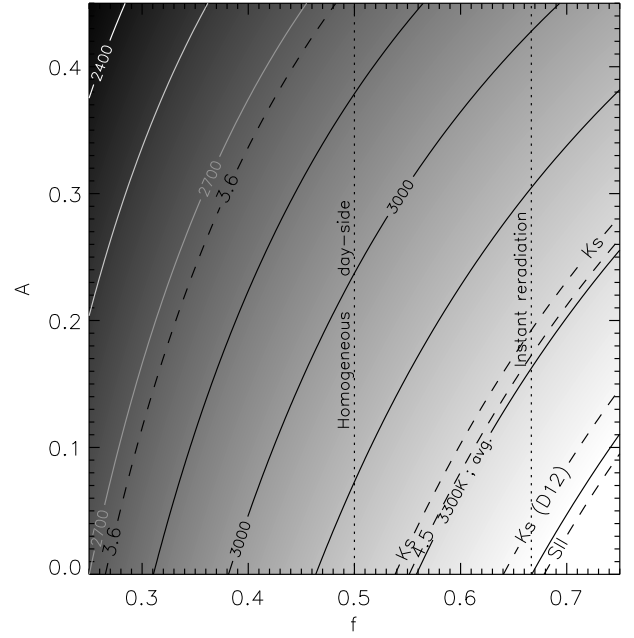}
\caption{Equilibrium temperature of \mbox{WASP-33b} for different albedo (A) and re-radiation factors (f). Solid contours show lines of constant temperatures at 150~K intervals. Overplotted are lines of constant temperature for the measured brightness temperatures (dashed lines), labeled with the bandpass they were observed (D12 indicates the K$_s$-band measurement from \cite{demingetal12}. The line labeled ``Avg.'' indicates the line for constant temperature of the effective temperature determined from the literature measurements. Vertical (dashed) lines indicate the re-radiation factors for a homogeneous day-side temperature (f=1/2), and for an instantly re-radiating day-side (f=2/3).}
\label{fig:fagrid}
\end{figure}

\subsection{A low albedo and rapid re-radiation of incident light}
With the exception of the {\it Spitzer} 3.6$\mu$m measurement from \cite{demingetal12}, all of the currently available eclipse measurements for \mbox{WASP-33b} point towards a very hot day-side temperature. If we assume that the measured brightness temperatures are representative of \mbox{WASP-33b}'s equilibrium temperature, and are not generated deep inside the planets atmosphere, where the temperatures are even higher, we can constrain the planet's equilibrium temperature to T$_{eff,p}$=3298$^{+66}_{-67}$K (see also Fig.~\ref{fig:sed}). Note that the basic assumption that the brightness temperature equals the effective temperature does not necessarily has to be the case, but detailed modelling of the available measurements is beyond the scope of this paper. From this equilibrium temperature we can further constrain the re-radiation factor and albedo. For the re-radiation factor we used the f description, as used by \cite{seagerandlopezmorales07}:
\begin{eqnarray}
T_p=T_*\left(\frac{R_*}{a}\right)^{1/2}[f(1-A_B)]^{1/4}
\end{eqnarray}
With R$_*$ the stellar radius, a the semi-major axis, A$_B$ the bond albedo and 1/4 $<$ f $<$ 2/3, where f=1/4 is for a homogenous temperature distriubution across the planet and f=2/3 is for instant re-radiation.

In Fig.~\ref{fig:fagrid} we show a simple model of the equilibrium temperature as a function of albedo and re-radiation factor. In addition we show lines of constant effective dayside temperature with the observed brightness temperatures and derived effective temperature overplotted. As can be seen the measurements require a very low albedo and a very short re-radiation time scale, such that all the stellar flux is absorbed and rapidly re-radiated without having time to advect to the night-side of the planet. This is consistent with the findings of \cite{cowanandagol11}, who study the albedo and redistribution efficiencies for a large sample of hot Jupiters, and find that the hottest planets (in their sample) have a low albedo and a low efficiency of the advection of absorbed stellar flux to the planet's night side.

The low redistribution efficiency suggests that the re-radiation timescales are short, and that the planet probably has an inversion layer \citep{fortneyetal08}. \cite{knutsonetal10} hypothesise that an increase in the UV-flux from an active star can cause a shift in the photochemistry such that the efficient absorber is removed from the gas-phase. The high incident UV-flux on \mbox{WASP-33b} would argue against this. It should be noted that for active stars most of the UV flux is emitted in the Lyman $\alpha$ line, while for WASP-33 it is likely that the UV continuum emission dominates. To investigate the influence of the UV-radiation, photochemical modelling will be necessary \citep[e.g.][]{zahnleetal09}.

\section{Conclusion}\label{sec:concl}
We have presented our results of K$_s$-band observations of the secondary eclipse of \mbox{WASP-33b}, the most irradiated planet known to date.  The measured eclipse depth for the second night is \depthsep\%, which results in a brightness temperature of \TBsep K. This high brightness temperature, if representative for the planet's equilibrium temperature, requires a very low albedo and a high (f$\gtrsim$0.5) re-radiation factor.

Combining our K$_s$-band measurement with the measurement of \cite{smithetal11}, we can fit a simple blackbody function to the spectral energy distribution, and determine an effective temperature of T$_{\rm{eff,p}}$=3290$^{+66}_{-67}$K.

We also find that stellar pulsations of the $\delta$~Scuti host-star, WASP-33, appears to have switched modes between the two nights, which are located a month apart, and also differ from the measurements by \cite{herreroetal11}. We caution, however, that this could be due to systematic effects which could also have strong periodicities.

The measurements on the first night suffer from strong residual systematics and stellar pulsations that cannot be fully corrected due to the short out-of-eclipse baseline, and demonstrates the need for observing a target for as long as possible.

\begin{acknowledgements}
We are grateful to the staff of the WHT telescope for their assistance with these observations. The William Herschel Telescope is operated on the island of La Palma by the Isaac Newton Group in the Spanish Observatorio del Roque de los Muchachos of the Instituto de Astrof\'isica de Canarias. 
\end{acknowledgements}

\bibliographystyle{aa} % style aa.bst
\bibliography{demooij_wasp33} % your references Yourfile.bib

\begin{thebibliography}{36}
\expandafter\ifx\csname natexlab\endcsname\relax\def\natexlab#1{#1}\fi

\bibitem[{{Acosta-Pulido} {et~al.}(2002){Acosta-Pulido}, {Ballesteros},
  {Barreto}, {Correa}, {Delgado}, {Dominguez-Tagle}, {Hernandez}, {Lopez},
  {Manchado}, {Manescau}, {Moreno}, {Prada}, {Redondo}, {Sanchez}, \&
  {Tenegi}}]{LIRIS02}
{Acosta-Pulido}, J., {Ballesteros}, E., {Barreto}, M., {et~al.} 2002, The
  Newsletter of the Isaac Newton Group of Telescopes, 6, 22

\bibitem[{{Alonso} {et~al.}(2010){Alonso}, {Deeg}, {Kabath}, \&
  {Rabus}}]{alonsoetal10}
{Alonso}, R., {Deeg}, H.~J., {Kabath}, P., \& {Rabus}, M. 2010, \aj, 139, 1481

\bibitem[{{Anderson} {et~al.}(2010){Anderson}, {Gillon}, {Maxted}, {Barman},
  {Collier Cameron}, {Hellier}, {Queloz}, {Smalley}, \&
  {Triaud}}]{andersonetal10}
{Anderson}, D.~R., {Gillon}, M., {Maxted}, P.~F.~L., {et~al.} 2010, \aap, 513,
  L3+

\bibitem[{{Breger}(1990)}]{breger90}
{Breger}, M. 1990, Delta Scuti Star Newsletter, 2, 13

\bibitem[{{Burrows} {et~al.}(2007){Burrows}, {Hubeny}, {Budaj}, {Knutson}, \&
  {Charbonneau}}]{burrowsetal07}
{Burrows}, A., {Hubeny}, I., {Budaj}, J., {Knutson}, H.~A., \& {Charbonneau},
  D. 2007, \apjl, 668, L171

\bibitem[{{C{\'a}ceres} {et~al.}(2011){C{\'a}ceres}, {Ivanov}, {Minniti},
  {Burrows}, {Selman}, {Melo}, {Naef}, {Mason}, \&
  {Pietrzynski}}]{caceresetal11}
{C{\'a}ceres}, C., {Ivanov}, V.~D., {Minniti}, D., {et~al.} 2011, \aap, 530, A5

\bibitem[{{Charbonneau} {et~al.}(2005){Charbonneau}, {Allen}, {Megeath},
  {Torres}, {Alonso}, {Brown}, {Gilliland}, {Latham}, {Mandushev}, {O'Donovan},
  \& {Sozzetti}}]{charbonneauetal05}
{Charbonneau}, D., {Allen}, L.~E., {Megeath}, S.~T., {et~al.} 2005, \apj, 626,
  523

\bibitem[{{Collier Cameron} {et~al.}(2010){Collier Cameron}, {Guenther},
  {Smalley}, {McDonald}, {Hebb}, {Andersen}, {Augusteijn}, {Barros}, {Brown},
  {Cochran}, {Endl}, {Fossey}, {Hartmann}, {Maxted}, {Pollacco}, {Skillen},
  {Telting}, {Waldmann}, \& {West}}]{colliercameronetal10}
{Collier Cameron}, A., {Guenther}, E., {Smalley}, B., {et~al.} 2010, \mnras,
  407, 507

\bibitem[{{Cowan} \& {Agol}(2011)}]{cowanandagol11}
{Cowan}, N.~B. \& {Agol}, E. 2011, \apj, 729, 54

\bibitem[{{Croll} {et~al.}(2010{\natexlab{a}}){Croll}, {Albert},
  {Lafreni{\`e}re}, {Jayawardhana}, \& {Fortney}}]{crolletal10a}
{Croll}, B., {Albert}, L., {Lafreni{\`e}re}, D., {Jayawardhana}, R., \&
  {Fortney}, J.~J. 2010{\natexlab{a}}, \apj, 717, 1084

\bibitem[{{Croll} {et~al.}(2010{\natexlab{b}}){Croll}, {Jayawardhana},
  {Fortney}, {Lafreni{\`e}re}, \& {Albert}}]{crolletal10b}
{Croll}, B., {Jayawardhana}, R., {Fortney}, J.~J., {Lafreni{\`e}re}, D., \&
  {Albert}, L. 2010{\natexlab{b}}, \apj, 718, 920

\bibitem[{{Croll} {et~al.}(2011){Croll}, {Lafreni{\`e}re}, {Albert},
  {Jayawardhana}, {Fortney}, \& {Murray}}]{crolletal10c}
{Croll}, B., {Lafreni{\`e}re}, D., {Albert}, L., {et~al.} 2011, \aj, 141, 30

\bibitem[{{de Mooij} {et~al.}(2012){de Mooij}, {Brogi}, {de Kok},
  {Koppenhoefer}, {Nefs}, {Snellen}, {Greiner}, {Hanse}, {Heinsbroek}, {Lee},
  \& {van der Werf}}]{demooijetal12}
{de Mooij}, E.~J.~W., {Brogi}, M., {de Kok}, R.~J., {et~al.} 2012, \aap, 538,
  A46

\bibitem[{{de Mooij} {et~al.}(2011){de Mooij}, {de Kok}, {Nefs}, \&
  {Snellen}}]{demooijetal11}
{de Mooij}, E.~J.~W., {de Kok}, R.~J., {Nefs}, S.~V., \& {Snellen}, I.~A.~G.
  2011, \aap, 528, A49+

\bibitem[{{de Mooij} \& {Snellen}(2009)}]{demooijandsnellen09}
{de Mooij}, E.~J.~W. \& {Snellen}, I.~A.~G. 2009, \aap, 493, L35

\bibitem[{{Deming}(2009)}]{deming09iau}
{Deming}, D. 2009, in Proceedings of the International Astronomical Union, Vol.
  253, , 197--207

\bibitem[{{Deming} {et~al.}(2012){Deming}, {Fraine}, {Sada}, {Madhusudhan},
  {Knutson}, {Harrington}, {Blecic}, {Nymeyer}, {Smith}, \&
  {Jackson}}]{demingetal12}
{Deming}, D., {Fraine}, J.~D., {Sada}, P.~V., {et~al.} 2012, \apj, 754, 106

\bibitem[{{Fortney} {et~al.}(2008){Fortney}, {Lodders}, {Marley}, \&
  {Freedman}}]{fortneyetal08}
{Fortney}, J.~J., {Lodders}, K., {Marley}, M.~S., \& {Freedman}, R.~S. 2008,
  \apj, 678, 1419

\bibitem[{{Fressin} {et~al.}(2010){Fressin}, {Knutson}, {Charbonneau},
  {O'Donovan}, {Burrows}, {Deming}, {Mandushev}, \& {Spiegel}}]{fressinetal10}
{Fressin}, F., {Knutson}, H.~A., {Charbonneau}, D., {et~al.} 2010, \apj, 711,
  374

\bibitem[{{Gelman} \& {Rubin}(1992)}]{gelmanandrubin92}
{Gelman}, A. \& {Rubin}, D.~B. 1992, Statistical Science, 7, 457

\bibitem[{{Gibson} {et~al.}(2010){Gibson}, {Aigrain}, {Pollacco}, {Barros},
  {Hebb}, {Hrudkov{\'a}}, {Simpson}, {Skillen}, \& {West}}]{gibsonetal10}
{Gibson}, N.~P., {Aigrain}, S., {Pollacco}, D.~L., {et~al.} 2010, \mnras, 404,
  L114

\bibitem[{{Gillon} {et~al.}(2007){Gillon}, {Demory}, {Barman}, {Bonfils},
  {Mazeh}, {Pont}, {Udry}, {Mayor}, \& {Queloz}}]{gillonetal07}
{Gillon}, M., {Demory}, B., {Barman}, T., {et~al.} 2007, \aap, 471, L51

\bibitem[{{Gillon} {et~al.}(2009){Gillon}, {Demory}, {Triaud}, {Barman},
  {Hebb}, {Montalb{\'a}n}, {Maxted}, {Queloz}, {Deleuil}, \&
  {Magain}}]{gillonetal09}
{Gillon}, M., {Demory}, B., {Triaud}, A.~H.~M.~J., {et~al.} 2009, \aap, 506,
  359

\bibitem[{{Handler} \& {Shobbrook}(2002)}]{handlerandshobbrook02}
{Handler}, G. \& {Shobbrook}, R.~R. 2002, \mnras, 333, 251

\bibitem[{{Hauschildt} {et~al.}(1999){Hauschildt}, {Allard}, {Ferguson},
  {Baron}, \& {Alexander}}]{hauschildtetal99}
{Hauschildt}, P.~H., {Allard}, F., {Ferguson}, J., {Baron}, E., \& {Alexander},
  D.~R. 1999, \apj, 525, 871

\bibitem[{{Herrero} {et~al.}(2011){Herrero}, {Morales}, {Ribas}, \&
  {Naves}}]{herreroetal11}
{Herrero}, E., {Morales}, J.~C., {Ribas}, I., \& {Naves}, R. 2011, \aap, 526,
  L10+

\bibitem[{{Knutson} {et~al.}(2010){Knutson}, {Howard}, \&
  {Isaacson}}]{knutsonetal10}
{Knutson}, H.~A., {Howard}, A.~W., \& {Isaacson}, H. 2010, \apj, 720, 1569

\bibitem[{{L{\'o}pez-Morales} {et~al.}(2010){L{\'o}pez-Morales}, {Coughlin},
  {Sing}, {Burrows}, {Apai}, {Rogers}, {Spiegel}, \&
  {Adams}}]{lopezmoralesetal10}
{L{\'o}pez-Morales}, M., {Coughlin}, J.~L., {Sing}, D.~K., {et~al.} 2010,
  \apjl, 716, L36

\bibitem[{{L{\'o}pez-Morales} \& {Seager}(2007)}]{seagerandlopezmorales07}
{L{\'o}pez-Morales}, M. \& {Seager}, S. 2007, \apjl, 667, L191

\bibitem[{{Madhusudhan} \& {Seager}(2010)}]{madhusudhanandseager10}
{Madhusudhan}, N. \& {Seager}, S. 2010, \apj, 725, 261

\bibitem[{{Mandel} \& {Agol}(2002)}]{mandelandagol02}
{Mandel}, K. \& {Agol}, E. 2002, \apjl, 580, L171

\bibitem[{{Rogers} {et~al.}(2009){Rogers}, {Apai}, {L{\'o}pez-Morales}, {Sing},
  \& {Burrows}}]{rogersetal09}
{Rogers}, J.~C., {Apai}, D., {L{\'o}pez-Morales}, M., {Sing}, D.~K., \&
  {Burrows}, A. 2009, \apj, 707, 1707

\bibitem[{{Sing} \& {L{\'o}pez-Morales}(2009)}]{singandlopezmorales09}
{Sing}, D.~K. \& {L{\'o}pez-Morales}, M. 2009, \aap, 493, L31

\bibitem[{{Smith} {et~al.}(2011){Smith}, {Anderson}, {Skillen}, {Collier
  Cameron}, \& {Smalley}}]{smithetal11}
{Smith}, A.~M.~S., {Anderson}, D.~R., {Skillen}, I., {Collier Cameron}, A., \&
  {Smalley}, B. 2011, \mnras, 416, 2096

\bibitem[{{Zahnle} {et~al.}(2009){Zahnle}, {Marley}, {Freedman}, {Lodders}, \&
  {Fortney}}]{zahnleetal09}
{Zahnle}, K., {Marley}, M.~S., {Freedman}, R.~S., {Lodders}, K., \& {Fortney},
  J.~J. 2009, \apjl, 701, L20

\bibitem[{{Zechmeister} \& {K{\"u}rster}(2009)}]{zechmeisterandkurster09}
{Zechmeister}, M. \& {K{\"u}rster}, M. 2009, \aap, 496, 577

\end{thebibliography}

\begin{appendix} 
\section{Correlation plots for the MCMC analysis}
\begin{figure*}
\centering
\includegraphics[width=16cm]{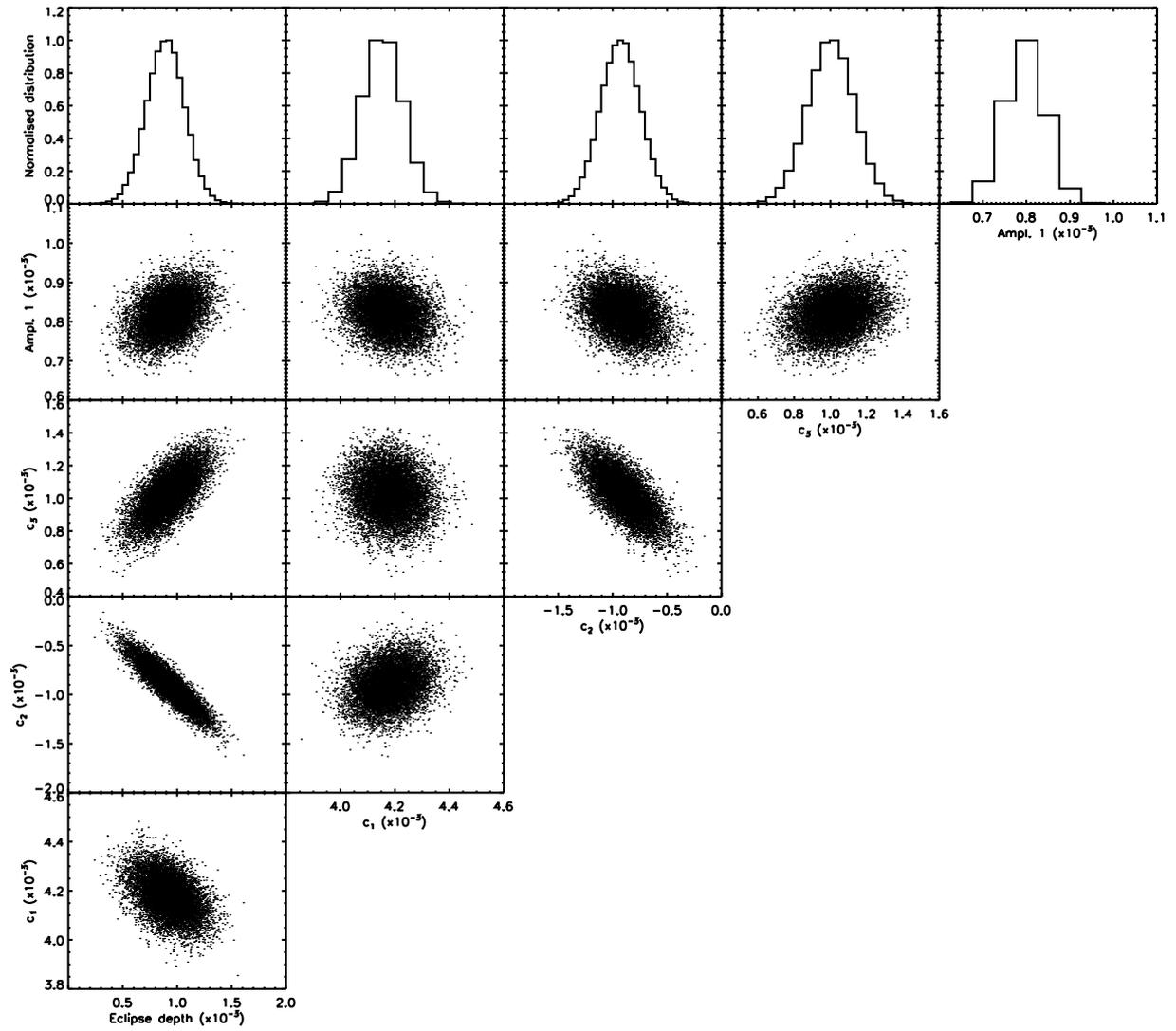}
\caption{Correlations between different parameters from the MCMC analysis for night I. The baseline is modelled using a polynomial with coefficients c$_1$ to c$_3$. The correlation between the parameters and the amplitudes of the stellar pulsation at lower significance are not shown, since their impact on any of the measured parameters will be minimal. The amplitude for stellar pulsation is for a pulsation period of 62.00 minutes.}
\label{fig:mcmc_cor_n1pol}
\end{figure*}
\begin{figure*}
\centering
\includegraphics[width=16cm]{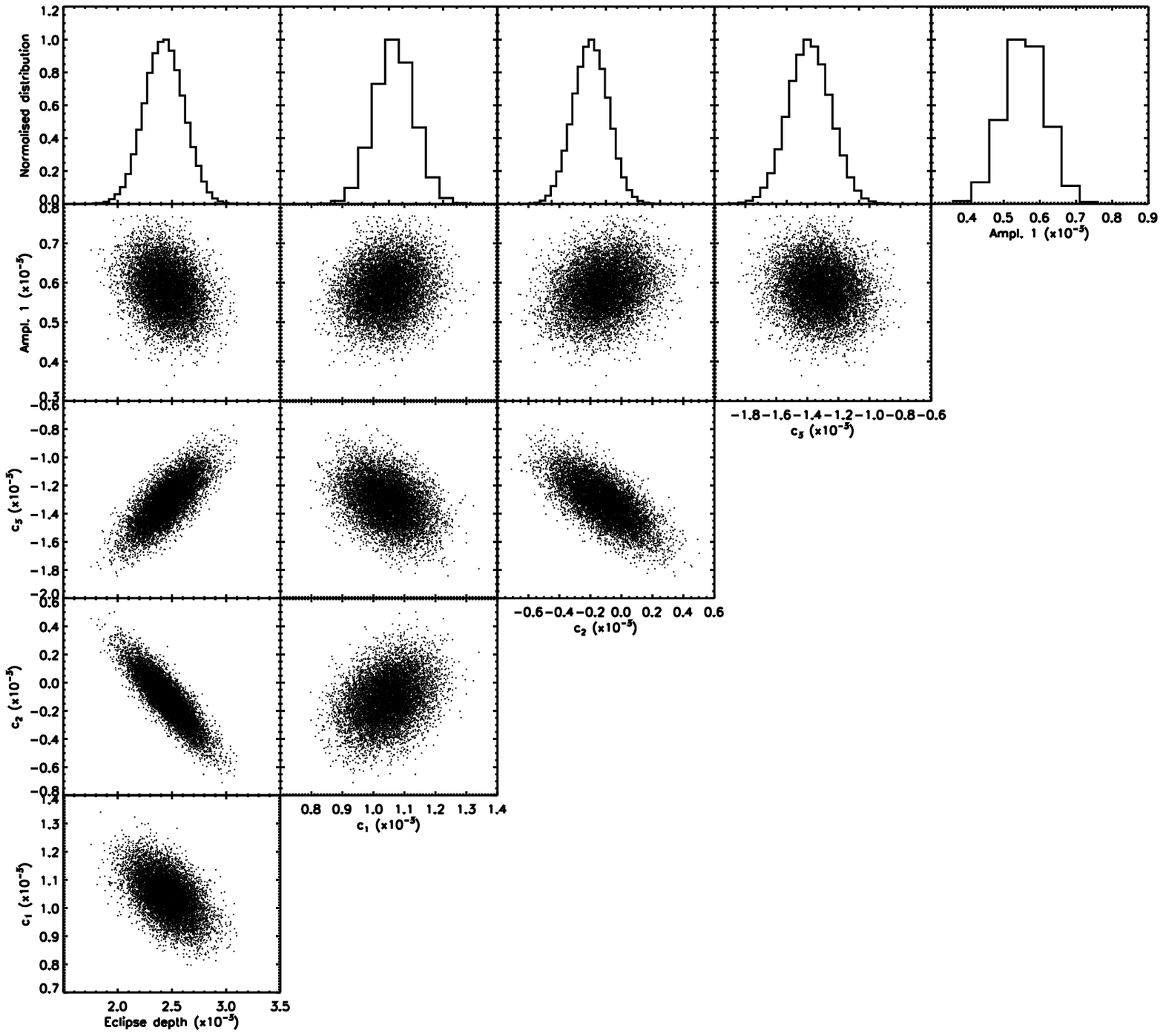}
\caption{same as Fig.~\ref{fig:mcmc_cor_n1pol} but now for night II. The amplitude for stellar pulsation is for a pulsation period of 52.28 minutes.}
\label{fig:mcmc_cor_n2pol}
\end{figure*}
\begin{figure*}
\centering
\includegraphics[width=16cm]{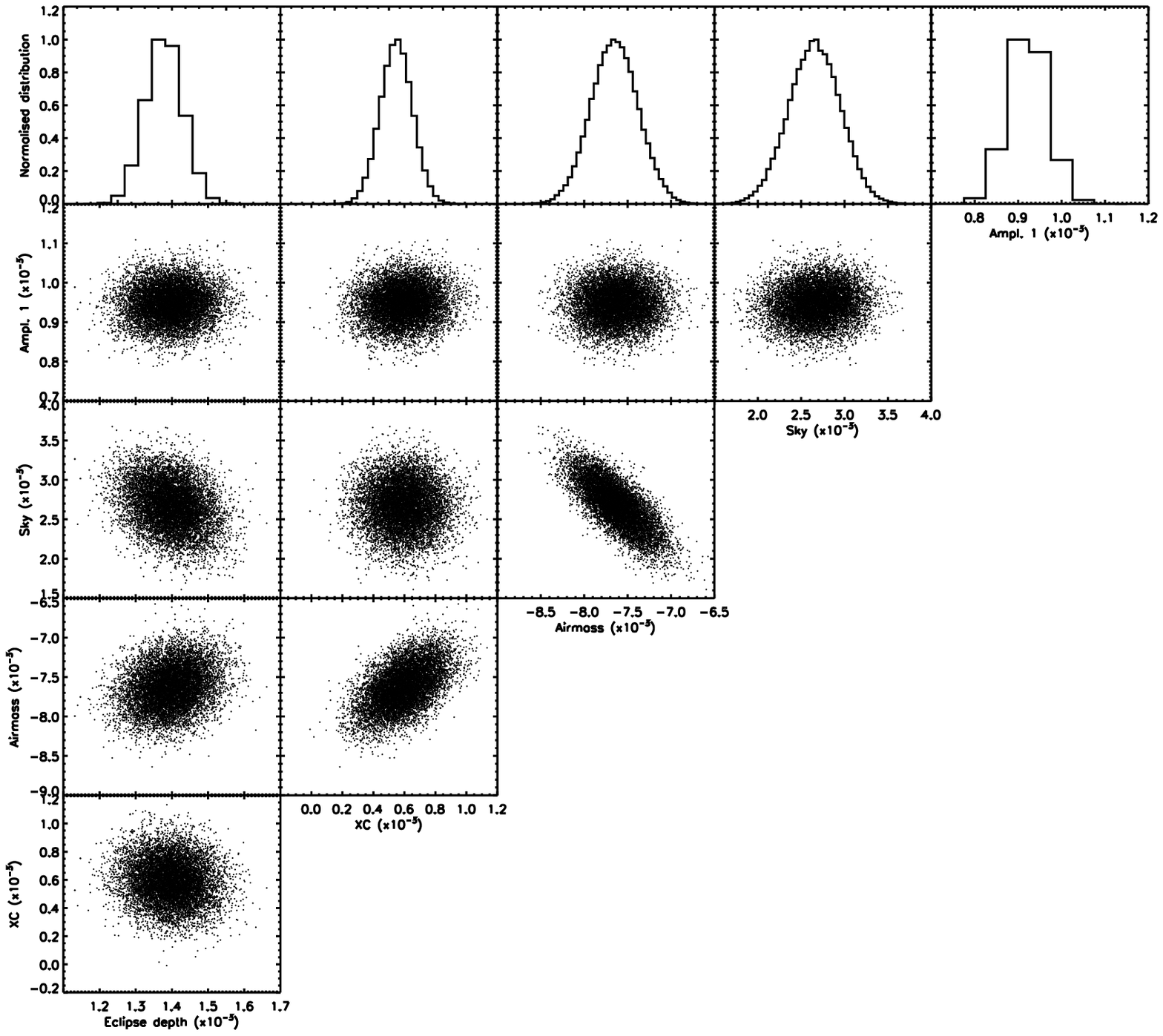}
\caption{Correlations between different parameters from the MCMC analysis for night I. The baseline is modelled using the observed instrumental parameters. The correlation between the parameters and the amplitudes of the stellar pulsation at lower significance are not shown, since their impact on any of the measured parameters will be minimal. The amplitude for stellar pulsation is for a pulsation period of 63.95 minutes.}
\label{fig:mcmc_cor_n1sys}
\end{figure*}
\begin{figure*}
\centering
\includegraphics[width=16cm]{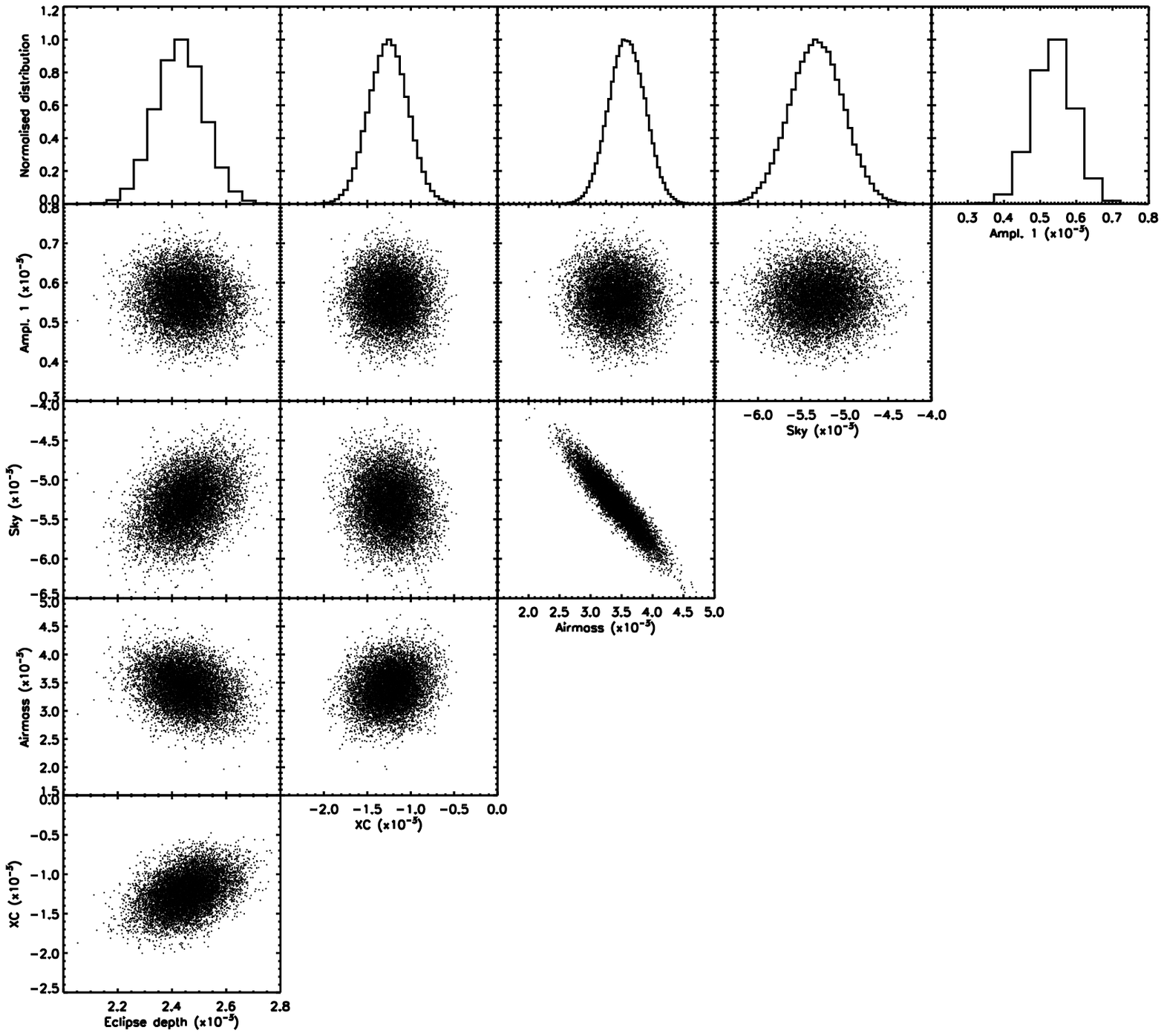}
\caption{same as Fig.~\ref{fig:mcmc_cor_n1sys} but for night II. The amplitude for stellar pulsation is for a pulsation period of 52.65 minutes.}
\label{fig:mcmc_cor_n2sys}
\end{figure*}
 
\end{appendix}

\end{document}